\documentclass[conference]{IEEEtran}
\IEEEoverridecommandlockouts

\usepackage{cite}
\usepackage{amsmath,amssymb,amsfonts}
\usepackage{graphicx}
\usepackage{textcomp}
\usepackage{xcolor}
\def\BibTeX{{\rm B\kern-.05em{\sc i\kern-.025em b}\kern-.08em
    T\kern-.1667em\lower.7ex\hbox{E}\kern-.125emX}}

\usepackage{subfig}
\usepackage{graphicx}
\usepackage{graphics}
\usepackage{epstopdf}
\usepackage{amsmath}
\usepackage{multirow}
\usepackage{bm}
\usepackage{enumerate}
\usepackage{mathdots}
\usepackage{todonotes}
\usepackage{algorithmicx}
\usepackage{algpseudocode}
\usepackage{algorithm}
\usepackage{verbatim}
\usepackage{balance}
\usepackage{cite}
\usepackage[hyphens]{url}
\usepackage{acro}

\usepackage{color}
\definecolor{purple(x11)}{rgb}{0.63, 0.36, 0.94}
\definecolor{cadmiumgreen}{rgb}{0.0, 0.42, 0.24}

\usepackage{amsmath,amssymb}

\newcommand{\bbC}{{\mathbb{C}}}

\newcommand{\bbE}{{\mathbb{E}}}

\newcommand{\bbR}{{\mathbb{R}}}

\newcommand{\ba}{{\mathbf{a}}}

\newcommand{\bd}{{\mathbf{d}}}

\newcommand{\bff}{{\mathbf{f}}}

\newcommand{\br}{{\mathbf{r}}}

\newcommand{\bu}{{\mathbf{u}}}

\newcommand{\bw}{{\mathbf{w}}}
\newcommand{\bx}{{\mathbf{x}}}

\newcommand{\bzero}{{\mathbf{0}}}

\newcommand{\bF}{{\mathbf{F}}}

\newcommand{\bH}{{\mathbf{H}}}
\newcommand{\bI}{{\mathbf{I}}}

\newcommand{\bR}{{\mathbf{R}}}

\newcommand{\bW}{{\mathbf{W}}}
\newcommand{\bX}{{\mathbf{X}}}
\newcommand{\bY}{{\mathbf{Y}}}

\newcommand{\rmc}{{\mathrm{c}}}

\newcommand{\rmr}{{\mathrm{r}}}

\newcommand{\rmF}{{\mathrm{F}}}

\newcommand{\cC}{\mathcal{C}}

\newcommand{\cN}{\mathcal{N}}

\newcommand{\bsfn}{\boldsymbol{\mathsf{n}}}

\newcommand{\bsfs}{\boldsymbol{\mathsf{s}}}

\newcommand{\bsfy}{\boldsymbol{\mathsf{y}}}

\newcommand{\sfT}{\mathsf{T}}

\newcommand{\bsfF}{\boldsymbol{\mathsf{F}}}

\newcommand{\bsfH}{\boldsymbol{\mathsf{H}}}

\newcommand{\bsfR}{\boldsymbol{\mathsf{R}}}

\newcommand{\bsfW}{\boldsymbol{\mathsf{W}}}

\newcommand{\transp}{{\sf{T}}}
\newcommand{\compj}{{\rm{j}}}

\makeatletter
\def\munderbar#1{\underline{\sbox\tw@{$#1$}\dp\tw@\z@\box\tw@}}
\makeatother

\DeclareAcronym{3GPP}{
  short=3GPP,
  long=3rd generation partnership project
}
\DeclareAcronym{ADC}{
  short=ADC,
  long=analog-to-digital converter
}
\DeclareAcronym{AMP}{
  short=AMP,
  long=approximate message passing
}
\DeclareAcronym{AoA}{
  short=AoA,
  long=angle-of-arrival
}
\DeclareAcronym{AoD}{
  short=AoD,
  long=angle-of-departure
}
\DeclareAcronym{APS}{
  short=APS,
  long=azimuth power spectrum
}
\DeclareAcronym{AV}{
  short=AV,
  long=autonomous vehicle
}
\DeclareAcronym{BS}{
  short=BS,
  long=base station
}
\DeclareAcronym{BSM}{
  short=BSM,
  long=basic safety message
}
\DeclareAcronym{CP}{
  short=CP,
  long=cyclic-prefix
}
\DeclareAcronym{DFT}{
  short=DFT,
  long=discrete Fourier transform
}
\DeclareAcronym{DL}{
  short=DL,
  long=downlink
}
\DeclareAcronym{DSRC}{
  short=DSRC,
  long=dedicated short-range communication
}
\DeclareAcronym{FDD}{
  short=FDD,
  long=frequency division duplex
}
\DeclareAcronym{FMCW}{
  short=FMCW,
  long=frequency modulated continuous wave
}
\DeclareAcronym{FoV}{
  short=FoV,
  long=field-of-view
}
\DeclareAcronym{GNSS}{
  short=GNSS,
  long=global navigation satellite system
}
\DeclareAcronym{LIDAR}{
  short=LIDAR,
  long=Light detection and ranging
}
\DeclareAcronym{LOS}{
  short=LOS,
  long=line-of-sight
}
\DeclareAcronym{LPF}{
  short=LPF,
  long=low pass filter
}
\DeclareAcronym{LTE}{
  short=LTE,
  long=long term evolution
}
\DeclareAcronym{MIMO}{
  short=MIMO,
  long=multiple-input multiple-output
}
\DeclareAcronym{MRR}{
  short=MRR,
  long=medium range radar
}
\DeclareAcronym{NLOS}{
  short=NLOS,
  long=non-line-of-sight
}
\DeclareAcronym{NR}{
  short=NR,
  long=new radio
}
\DeclareAcronym{OFDM}{
  short=OFDM,
  long=orthogonal frequency-division multiplexing
}
\DeclareAcronym{ppm}{
  short=ppm,
  long=parts-per-million
}
\DeclareAcronym{RMS}{
  short=RMS,
  long=root-mean-square
}
\DeclareAcronym{RPE}{
  short=RPE,
  long=relative precoding efficiency
}
\DeclareAcronym{RSU}{
  short=RSU,
  long=roadside unit
}
\DeclareAcronym{SNR}{
  short=SNR,
  long=signal-to-noise ratio
}
\DeclareAcronym{UL}{
  short=UL,
  long=uplink
}

\DeclareAcronym{ULA}{
  short=ULA,
  long=uniform linear array
}
\DeclareAcronym{V2I}{
  short=V2I,
  long=vehicle-to-infrastructure
}
\DeclareAcronym{V2V}{
  short=V2V,
  long=vehicle-to-vehicle
}
\DeclareAcronym{V2X}{
  short=V2X,
  long=vehicle-to-everything
}
\DeclareAcronym{VRU}{
  short=VRU,
  long=vulnerable road user
}

\usepackage{pifont}

\newcommand{\be}{\begin{eqnarray}}
\newcommand{\ee}{\end{eqnarray}}

\def\ba{{\mathbf{a}}}

\def\bd{{\mathbf{d}}}

\def\bff{{\mathbf{f}}}

\def\br{{\mathbf{r}}}

\def\bu{{\mathbf{u}}}

\def\bw{{\mathbf{w}}}
\def\bx{{\mathbf{x}}}

\def\b0{{\mathbf{0}}}

\def\bF{{\mathbf{F}}}

\def\bH{{\mathbf{H}}}
\def\bI{{\mathbf{I}}}

\def\bR{{\mathbf{R}}}

\def\bW{{\mathbf{W}}}
\def\bX{{\mathbf{X}}}
\def\bY{{\mathbf{Y}}}

\def\sfT{\mathsf{T}}

\def\bsfF{\bm{\mathsf{F}}}

\def\bsfH{\bm{\mathsf{H}}}

\def\bsfR{\bm{\mathsf{R}}}

\def\bsfW{\bm{\mathsf{W}}}

\def\bsfn{{\bm{\mathsf{n}}}}

\def\bsfs{{\bm{\mathsf{s}}}}

\def\bsfy{{\bm{\mathsf{y}}}}

\def\bsf0{{\bm{\mathsf{0}}}}

\newcommand{\Ns}{N_{\rm{s}}}

\newcommand{\NRSU}{N_{\rm{R}\rm{S}\rm{U}}}
\newcommand{\NV}{N_{\rm{V}}}
\newcommand{\MRSU}{M_{\rm{R}\rm{S}\rm{U}}}
\newcommand{\MV}{M_{\rm{V}}}
\newcommand{\bFBB}{\bF_{\rm{B}\rm{B}}}
\newcommand{\bFRF}{\bF_{\rm{R}\rm{F}}}
\newcommand{\bWBB}{\bW_{\rm{B}\rm{B}}}
\newcommand{\bWRF}{\bW_{\rm{R}\rm{F}}}
\newcommand{\alpharc}{\alpha_{r_c}}
\newcommand{\baRSU}{\ba_{\rm{R}\rm{S}\rm{U}}}
\newcommand{\baV}{\ba_{\rm{V}}}

\newcommand{\bsfRRSU}{\bsfR_{\rm{R}\rm{S}\rm{U}}}
\newcommand{\hbsfRRSU}{\hat{\bsfR}_{\rm{R}\rm{S}\rm{U}}}

\newcommand{\Tr}{T_{\rm{r}}}
\newcommand{\Tc}{T_{\rm{c}}}

\newcommand{\VN}[1]{\textbf{#1}}

\newcommand{\degree}{^{\circ}}

\begin{document}
\title{Radar  Aided mmWave Vehicle-to-Infrastructure Link Configuration Using Deep Learning}
\author{Yun Chen$^{\dag}$, Andrew Graff$^\ddag$, Nuria Gonz\'{a}lez-Prelcic$^{\dag}$  and Takayuki Shimizu$^\S$\thanks{ This work has been partially funded by Toyota Motor North America.} \\
	$^{\dag}$ North Carolina State University, Email: \{ychen273, ngprelcic\}@ncsu.edu \\
	$^\ddag$ The University of Texas at Austin, Email: andrewgraff@utexas.edu\\
	$^\S$ Toyota Motor North America, Email: takayuki.shimizu@toyota.com} 

\maketitle

\begin{abstract}
The high overhead of the beam training process is the main challenge when establishing mmWave communication links, especially for vehicle-to-everything (V2X) scenarios where the channels are highly dynamic.
In this paper, we obtain prior information to speed up the beam training process by implementing two deep neural networks (DNNs) that  realize radar-to-communication (R2C) channel information translation in a vehicle-to-infrastructure (V2I) system. Specifically, the first DNN is built to extract the information from the radar azimuth power spectrum (APS) to reconstruct the communication APS, while the second DNN exploits the information extracted from the spatial covariance of the radar channel to realize R2C covariance prediction. 
The achieved data rate and the similarity between the estimated and the true communication APS are used to evaluate the prediction performance. The covariance estimation method generally provides  higher similarity, as the APS predictions cannot always capture the mismatch between the radar and communication APS. Compared to the beam training method which exploits directly the radar APS without an attempt to translate it to the communication channel, our proposed deep learning (DL) aided methods remarkably reduce the beam training overhead, resulting in a 13.3\% and 21.9\% rate increase when using the communication APS prediction and covariance prediction, respectively.
\end{abstract}

%

\section{Introduction}
MmWave technology has been deployed for large spectrum demands of the fifth-generation (5G) mobile networks. 
To support high throughput data transfer in 5G mmWave MIMO systems, beamforming 
focuses the wireless signals towards specific directions to compensate for the small antenna aperture and achieve satisfying signal-to-noise ratios (SNR) at the receiver (RX). However,  when large antenna arrays are deployed at both the transmitter (TX) and RX, the overhead of the standard beam-training process is generally high. 
The problem is especially pertinent to highly mobile scenarios, e.g., V2X communication, where the mmWave-vehicular channels are highly dynamic.

Efforts have been made in prior work to reduce the overhead of the beam training process by exploiting out-of-band information,  
which can be  related to mmWave link information. For example, spatial information at sub-6 GHz can be extracted to help establish the mmWave link by solving a weighted sparse signal recovery problem \cite{ali2017millimeter} or for spatial covariance translation \cite{ali2019spatial}. Previous works also propose sensor-aided beam training methods. Thus, localization information obtained through radars and automotive sensors is capable of assisting beam alignment \cite{garcia2016location, va2019online}. Prior location information can also used to speed up the adaptive channel estimation and beamforming stage \cite{garcia2016location}. An online learning algorithm for position-aided beam pair selection and refinement was proposed in \cite{va2019online}. LIDAR data can also be used for beam selection based on deep learning as proposed in \cite{klautau2019lidar}. Finally, radar covariance information has also been used in previous work \cite{anum2020passive,Prelcic2016} as a direct estimate of the communication covariance to reduce beamtraining overhead.

Work described above has, however, some limitations. 
The strategies proposed in \cite{ali2017millimeter, ali2019spatial} require the state of the sub-6 GHz and the mmWave link to be strictly the same (both are either line-of-sight (LOS) or non-line-of-sight (NLOS)), while the location-aided search \cite{garcia2016location} and the LIDAR aided beam selection \cite{klautau2019lidar} only reduce the effective beam search areas in the presence of LOS propagation. Though the online-learning algorithm in \cite{va2019online} works in NLOS, the time complexity is compromised using the algorithm with lots of iterations. Regarding the previous approaches that exploits radar covariance information \cite{anum2020passive,Prelcic2016}, the main limitation is the inherent mismatch between the true communication covariance and the radar communication covariance used as its estimate. 

DL is promising for translating radar data to useful information about the communication channel, including mismatches between the propagation scenarios for radar and communication. In this paper, we present two DNN architectures for vehicle-to-infrastructure (V2I) communication link configuration that provide a mapping between the radar and the communication spatial  information.  
Our proposed methods are applicable for NLOS scenarios, and we provide system-level simulation results to validate our models. 
The results show that the APS prediction method allows for 181 Mbps (13.3\%) rate increase compared  to the beam training method exploiting directly the radar APS as a prior without learning the mismatch (radar-only search), while the covariance prediction method achieves a 12.4\% higher rate than the APS prediction strategy.

{\em Notation:} 
$x$, $\VN{x}$, $\VN{X}$ represent a scalar, a column vector and a matrix, respectively; $\bx_r$, $\bX_r$ denote the vector and matrix related to the passive radar channel, and $\bx_c$, $\bX_c$ are related to the communication channel. 

 \section{System model}

Our model is based on the \ac{V2I} communication system described in ~\cite{anum2020passive}. In this system, we have a \ac{RSU} on the side of a roadway and an ego vehicle driving along the road. The \ac{RSU} is equipped with a passive radar \ac{ULA} and a communications \ac{ULA}. The ego vehicle has 4 \ac{ULA}s for communications and 4 single-antenna automotive radars. The communication arrays are placed in accordance with \ac{3GPP} proposals~\cite{3GPP37885}, and the radar arrays are placed at the front and rear corners to model the mid range radar deployment of a Lexus LS or a Toyota Mirai \cite{Toyota2020}. 
The passive radar array at the \ac{RSU} will use receptions of the automotive radar signals to estimate the radar spatial covariance. This covariance will then be used to configure the mmWave communication link.

\subsection{Communication system model}\label{sec:comm}

The communication array on the \ac{RSU} is equipped with $\NRSU$ antennas and $\MRSU\leq\NRSU$ RF-chains. We let $A=4$ denote the number of communication arrays at the ego vehicle. Each vehicle array has $\NV$ antenna elements and $\MV\leq\NV$ RF-chains. This hybrid architecture supports $\Ns\leq\min\{\MRSU,\MV\}$ data-streams. The communication link is a $K$ subcarrier \ac{OFDM} system, with modulated symbols $\bsfs[k]\in\bbC^{\Ns\times1}$ such that $\bbE[\bsfs[k]\bsfs^\ast[k]]=\frac{P_\rmc}{K\Ns}\bI_{\Ns}$ and  $P_\rmc$ denotes the total average transmitted power. The baseband precoder $\bFBB[k]\in\bbC^{\MRSU\times\Ns}$ and RF precoder $\bFRF\in\bbC^{\NRSU\times\MRSU}$ are combined to form the hybrid precoder $\bF[k]=\bFRF\bFBB[k]\in\bbC^{\NRSU\times\Ns}$ on subcarrier $k$. The RF precoder is realized using quantized phase shifters and is the same across all subcarriers. Letting $\zeta_{i,j}$ be the quantized phase, quantization is described as $[\bFRF]_{i,j}=\frac{1}{\sqrt{\NRSU}}e^{\compj \zeta_{i,j}}$. The total power constraint is enforced as $\sum_{k=1}^{K}\|\bF[k]\|_\rmF^2=K\Ns$. 

The baseband combiner $\bWBB^{(a)}[k]\in\bbC^{\MV\times\Ns}$ and RF combiner $\bWRF^{(a)}\in\bbC^{\NV\times\MV}$ are combined to form the hybrid combiner $\bW^{(a)}[k]=\bWRF^{(a)}\bWBB^{(a)}[k]\in\bbC^{\NV\times\Ns}$ on subcarrier $k$. The $\NV\times\NRSU$ frequency-domain MIMO channel at array $a\in{A}$ is denoted as $\bsfH^{(a)}[k]$. Assuming perfect synchronization, the received signal on subcarrier $k$ after processing is
\begin{align}
\bsfy^{(a)}[k]=\bsfW^{(a)\ast}[k]\bsfH^{(a)}[k]\bsfF[k]\bsfs[k]+\bsfW^{(a)\ast}[k]\bsfn^{(a)}[k],
\label{eq:rxpost}
\end{align}
where $\bsfn^{(a)}\sim\cC\cN(\bzero,\sigma_{\bsfn}^2\bI)$ is additive white Gaussian noise.

\subsection{Channel model} 

The wideband channel is modeled geometrically with $C$ clusters. Each of the clusters experiences a mean time delay $\tau_c \in \bbR$, mean \ac{AoA} $\theta_c \in [0,2\pi)$, and mean \ac{AoD} $\phi_c \in [0,2\pi)$. Assuming there are $R_c$ paths in each cluster, each path $r_c\in[R_c]$ has complex gain $\alpha_{r_c}$, relative time-delay $\tau_{r_c}$, relative arrival angle shift $\vartheta_{r_c}$, and relative departure angle shift $\varphi_{r_c}$. The array response vectors are $\baRSU(\phi)$ at the \ac{RSU} and $\baV(\theta)$ at the ego-vehicle. The uniform spacing between array elements is $\Delta$, normalized to units of wavelength. The \ac{RSU} response vector and ego-vehicle response vectors are defined as
\begin{align}
	\baRSU(\theta)=[1,e^{\compj 2\pi\Delta\sin(\theta)},\cdots,e^{\compj(\NRSU-1) 2\pi\Delta\sin(\theta)}]^\transp.
\end{align}
\begin{align}
	\baV(\phi)=[1,e^{\compj 2\pi\Delta\sin(\phi)},\cdots,e^{\compj(\NV-1) 2\pi\Delta\sin(\phi)}]^\transp.
\end{align}
We will remove the notation $(a)$ in the channel $\bH$ for the following equations. We will define the analog filtering and pulse shaping effect at delay $\tau$ as $p(\tau)$. $\Tc$ will denote the signaling interval. The delay-$d$ \ac{MIMO} channel matrix $\bH[d]$ is~\cite{anum2020passive}
\begin{align}
	\bH[d]=\sum_{c=1}^C \sum_{r_c=1}^{R_c} \alpharc p& (d\Tc-\tau_c-\tau_{r_c})\times\nonumber\\
	&\baV(\phi_c+\varphi_{r_c})\baRSU^\ast(\theta_c+\vartheta_{r_c}).
	\label{eq:timedomch}
\end{align}
If there are $D$ delay-taps in the channel, the channel at subcarrier $k$, $\bsfH[k]$ is~\cite{anum2020passive}
\begin{align}
	\bsfH[k]=\sum_{d=0}^{D-1}\bH[d] e^{-\compj \tfrac{2\pi k}{K}d}.
	\label{eq:freqdomch}
\end{align}

\subsection{Covariance model}\label{sec:covmod}
We define the spatial covariance at the \ac{RSU} on subcarrier $k$ as $\bsfRRSU[k]=\frac{1}{\NV}\bbE[\bsfH^\ast[k]\bsfH[k]]$. By assuming that the covariance does not change across subcarriers, we can create an estimate by averaging over all subcarriers $\hbsfRRSU=\frac{1}{K}\sum_{k=1}^K \hbsfRRSU [k]$. Since we will only use covariance estimates to design the analog precoder and combiner, this is an appropriate assumption, as the baseband precoder and combiner will be designed independently and account for subcarrier-dependent covariance variations~\cite{anum2020passive}.

\subsection{Radar system model}\label{sec:rad}
Each of the radars on the ego-vehicle transmit an arbitrary baseband signal $s_\rmr(t)$ modulated to a carrier frequency of $f_\rmr$. This transmission is scaled to achieve a transmit power $P_\rmr$, denoted as
\begin{align}    
s(t)&=\sqrt{P_\rmr}s_\rmr(t).
\end{align}
The received signal on the $N_\rmr$ element antenna array on the \ac{RSU} will be denoted as a vector $\bx(t)\in \bbC^{N_\rmr}$. Experiencing an attenuation of $\alpha$ and a time delay of $\tau_n$ during propagation to the $n$th antenna, the received signal at antenna $n$ is
\begin{align}
[\bx(t)]_n=\alpha s(t-\tau_n).
\end{align}

We can model the propagation delay as the sum of two components: one accounting for common distance $\tau$ and another accounting for the difference among antenna elements at the \ac{ULA} $\tau^\prime_n$. This delay at antenna $n$ is described as $\tau_n=\tau + \tau^\prime_n$~\cite{Katkovnik2002High}. We assume our \ac{ULA} has half-wavelength spacing, so $\tau^\prime_n=\frac{\sin \theta (n-1) }{2 f_\rmr}$.

Then, we collect the $I$ samples of the signal into a matrix $\bY\in\bbC^{N_\rmr \times I}$. Let $i\in\{1,2,\cdots,I\}$ denote the sample index, and $\Tr$ denote the sampling time. With this notation, the $i$th sample on the $n$th antenna can be written as
$ [\bY]_{n,i}=[\bx(i \Tr)]_n$, and the spatial covariance of the received radar signal is estimated as
\begin{align}
\bR=\frac{1}{I}\bY\bY^\ast.
\label{eq:perfcorrdef}
\end{align}
To simulate ideal covariance estimation, we let $s_\rmr(t) = 1$, creating a pure sinusoid at our carrier frequency. Other waveforms such as FMCW will introduce an angular bias into the covariance estimate, but this can be corrected~\cite{anum2020passive}.

\section{Radar-to-Communication APS and Covariance Prediction Using DL}\label{algs}

\subsection{Background and Motivation}
As the radar and communication channels share the same propagation environment, we aim to configure communication links based on the radar channel information to reduce the beam-training overhead. Previous work \cite{Prelcic2016,anum2020passive} directly uses the radar covariance information as an estimate of the communication covariance, without any attempt to translate it to the communication channel, or in other words, without any strategy to reduce the mismatch 
between the communication and the radar channel. This inherent mismatch is due to the different operation frequencies, different antenna geometry and size for the communication and radar systems, and different location of the communication and radar antenna arrays, both at the BS and the vehicle.

Developing an analytical model of the mismatch between radar and communication covariances is a challenging problem, especially in NLOS scenarios, due to the large amount of parameters that impact the mismatch.  Under these circumstances, we will show that a DNN is an architecture that can learn the mismatch between radar and communication spatial covariances from the data stored in an appropriate dataset, without need of exploiting any complex mathematical model.

We propose and describe two DNNs, the first one for R2C APS prediction, and the second one for R2C covariance prediction. We design the DNNs separately because the APS and the spatial covariance matrix contain a different representation of the channel information, and it is important to capture the local features like sharp peaks of the APS for APS prediction, while there are barely obvious local spacial coherence for the covariance matrix. 

\subsection{R2C APS Prediction}

The APS gives the distribution of power as a function of the azimuth angle. The beam search space can be reduced by referring to the angles corresponding to the peaks in the communication APS. In this work, we first acquire both the radar and communication spatial covariance matrices and then extract the APS from the covariance matrix \cite{anum2020passive}. Specifically, we first construct a DFT matrix $\bF=[\bff(\theta_1),...,\bff(\theta_N)]$, where $\bff(\theta_i)=[e^{-j\cdot 0\cdot 2\pi \Delta\sin(\theta_i)},..., e^{-j\cdot (N_{RSU}-1)\cdot 2\pi \Delta\sin(\theta_i)}]^{\sfT}$ and $\theta_i$ is uniformly spaced from $-\frac{\pi}{2}$ to $\frac{\pi}{2}$. Then, the APS can be obtained by extracting the diagonal elements of $\bF^*\bR\bF$, i.e., $\bd = diag(\bF^*\bR\bF)$. A DNN is designed to predict the communication APS $\hat{\bd}_c$ based on the radar APS $\bd_r$ to assist beam search. As the APS has obvious local features and we can observe a ``shape" of the power distribution, the DNN for APS prediction is mainly based on 1-D convolutional layers for feature extraction. In this work, we use $N=N_\text{RSU}$, which satisfies the Nyquist sampling of the array response, and accordingly $\bd_r (\bd_c) \in \mathbb{R}^{1\times N_\text{RSU}}$.

We design a DNN, denoted as $\mathcal{N}_\text{APS}(\cdot)$, for R2C APS prediction, which aims to approximate the true communication APS with the known radar APS, i.e.,
\begin{equation}
	\hat{\bd}_c = \mathcal{N}_\text{APS}(\bd_r;\bw),
\end{equation}
where $\bd_r$ and $\hat{\bd}_c$ are the given radar APS and predicted communication APS, and $\bw$ represents the network parameters of $\mathcal{N}_\text{APS}(\cdot)$ to be trained.

\begin{figure*}
	\centering
	\includegraphics[width=\textwidth]{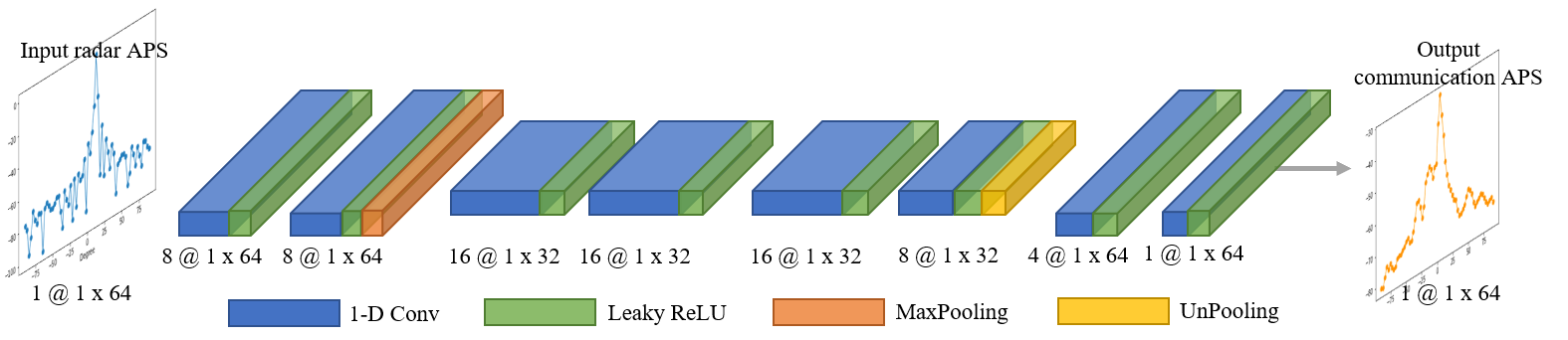}
	\caption{An illustration of the architecture of $\mathcal{N}_{\text{APS}}$. It is only convolutional without fully connected layers.}
	\label{APS_arch}
\end{figure*}

As the APS is the power distribution over all the directions of arrival, we can observe peaks on the direction where the energy is highly concentrated. This is a kind of local feature that we want to capture when linking the radar and communication APS through $\mathcal{N}_\text{APS}(\cdot)$, based on the fact that radar and communication channels share the same propagation environment. We apply 1-D convolutional layers in the network, since convolutional layers are well suited for capturing local features. In addition, as the components in the APS are real and non-negative values, we use LeakyReLU \cite{xu2015empirical} as the activation function in the network, which helps to avoid both dying ReLU and vanishing gradient problems during the training process. The detailed architecture of the network is shown in Fig.~\ref{APS_arch}. It is like an Encoder-Decoder network \cite{badrinarayanan2017segnet} where the MaxPooling layers reduce the input dimension and extract the local features of the radar APS efficiently, while the UpPooling layers are used to reconstruct the communication APS based on the extracted features. As the target for the prediction is to approximate the true communication APS for beam training, the mean squared error (MSE) between the predicted communication APS and the true one is considered as the loss function. The loss on the validation dataset is monitored during the training process for the purpose of early stopping against overfitting.

\subsection{R2C Covariance Column Prediction}

The spatial covariance matrix contains redundant information and it can be characterized by a featured column after Toeplitz completion \cite{anum2020passive}. The idea is to project the measured covariance matrix to a Toeplitz, Hermitian and positive semi-definite cone. Thus, the projected matrix $\widetilde{\bR}$ can be fully described by its first column $\br \in\mathbb{C}^{N_\text{RSU}\times 1}$. The DNN designed for covariance  prediction outputs the estimation for the column of the communication channel $\hat{\br}_c$ when the input is the column of the radar covariance $\br_r$. Considering that the column is structure agnostic (no special assumption needs to be made about the input), we choose fully connected (FC) layers as the basis for the network. The real and imaginary parts of the column are treated as a 2-channel input for the network, and the output is also 2-channel containing the real and imaginary parts of the predicted communication column.

\begin{figure}
	\centering
	\includegraphics[width=.5\textwidth]{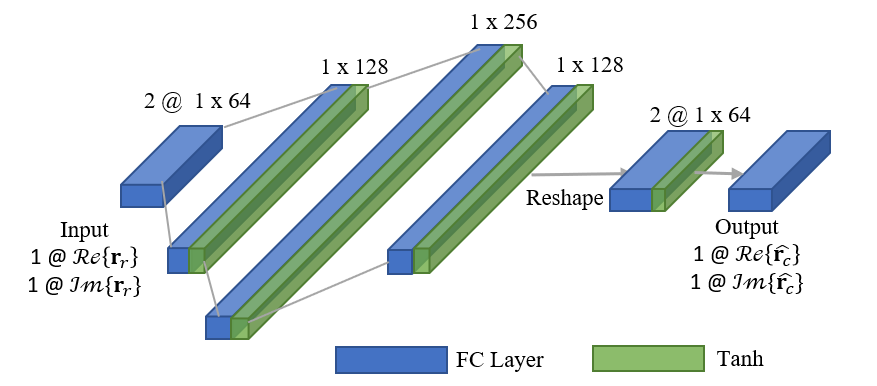}
	\caption{An illustration of the architecture of $\mathcal{N}_{\text{col}}$. }
	\label{Col_arch}
\end{figure}

A DNN denoted as $\mathcal{N}_\text{col}(\cdot)$ is designed for R2C covariance column prediction. Thus, our proposed covariance column prediction can be described as 
\begin{equation}\label{col_pred_equ}
	\hat{\br}_c = \mathcal{N}_{\text{col}}(\br_r;\bu),
\end{equation}
where $\bu$ contains the network parameters to be trained. The structure of $\mathcal{N}_\text{col}(\cdot)$ is different from $\mathcal{N}_{\text{APS}}(\cdot)$. First, the covariance column contains both real and imaginary parts, thus we treat the two parts of $\br_r$ as a 2-channel input for the network. Second, as the covariance column is not like the APS, which has obvious local spatial coherence, the network is mainly built with FC layers, which offer learning features from all the combinations of the features embedded in the input. There are three hidden FC layers in the network, which are sufficient considering the complexity of our prediction problem while making the DL process computational-friendly. Tanh is selected as the activation function, which constrains the values to $[-1,1]$, since the input can be either positive or non-positive values. The network architecture is shown in Fig.~\ref{Col_arch}, where the 2-channel output contains the real and imaginary parts of $\hat{\br}_c$. 

The loss function $\mathcal{L}({\rm \bu})$ is computed by transforming the column back to the Toeplitz matrix $\widetilde{\bR}(\br)$, calculating the APS, and evaluating the MSE between the predicted and true APS, i.e.,
\begin{equation}
	\mathcal{L}({\rm \bu})=\mathbb{E}\{||\bF^*\widetilde{\bR}(\hat{\br}_c)\bF-\bd_c||^2\}.
\end{equation}
The training process minimizes the loss on the training dataset, and we monitor the loss on the validation dataset for early stopping.


\section{Results}\label{sec:results}
In this section, we introduce the experimental setting and present the numerical results on the performance of our radar-aided communication link configuration using two different DNNs. First, the DL prediction accuracy is evaluated in Section \ref{simi_results} using the similarity metric for the APS defined in \cite{anum2020passive}. Then, the rate results are provided in Section \ref{sec_rate_results} to verify that our methods reduce the beam training overhead and achieve a rate increase when compared to the radar-aided beamtraining method  defined in previous work\cite{anum2020passive}, that ignores the mismatch between radar and communication channels.


\subsection{Dataset Generation}
We use Wireless Insite to conduct ray-tracing simulations to create the training and testing dataset. In the simulations, the RSU has $N_{RSU}=64$ antennas in the communication array, 64 antennas in the radar array, and the radar and communication arrays are horizontally aligned and are vertically separated by 10 cm. There are $N_v=16$ antennas per communication array on the vehicles. The ULAs used in both the communication and radar systems have half-wavelength inter-element spacing. The RSU arrays are down-tilted and communicate with vehicles in a 60 m length section of the roadway.  The communication system operates in the 73 GHz band with a bandwidth $B_c=1$ GHz. We choose the number of subcarriers $K=2048$ and the transmit power $P_c=24$ dBm. The raised-cosine filter with a roll-off factor 0.4 is used for pulse shaping. Accordingly, the number of time-domain taps required can be calculated to be 512. The radar system operates in the 76 GHz band. The transmit power $P_r$ is 30 dBm and the bandwidth $B_r$ is 1 GHz. Other parameters regarding the vehicles and the urban environment follow the deployment in \cite{anum2020passive}. 

Each ray-tracing simulation outputs a communication channel, a communication covariance matrix measured at the \ac{RSU}, and a radar covariance matrix also measured at the RSU. The dataset was created from 16000 of these simulations. These were randomly divided into a training set of 9600 entries, a validation set of 2400 entries, and a test set of 4000 entries. 

\subsection{DL Simulation Settings}
The APS is usually sparse, so we transform the APS to a logarithmic scale. This way, the neurons in the DNN for APS prediction could be effectively activated for accurate approximations. For the covariance column prediction, we simply normalize the dataset to [-1,1]. For each one of the two methods, the training batch size is set to 64, and Adam \cite{kingma2014adam} is used as the optimizer with the learning rate set to 0.001. The training process for each method takes 1000 iterations with early stopping. 
\subsection{Similarity Performance of the DNN Predictions}\label{simi_results}
\begin{figure*}
	\centering
	\includegraphics[width=\textwidth]{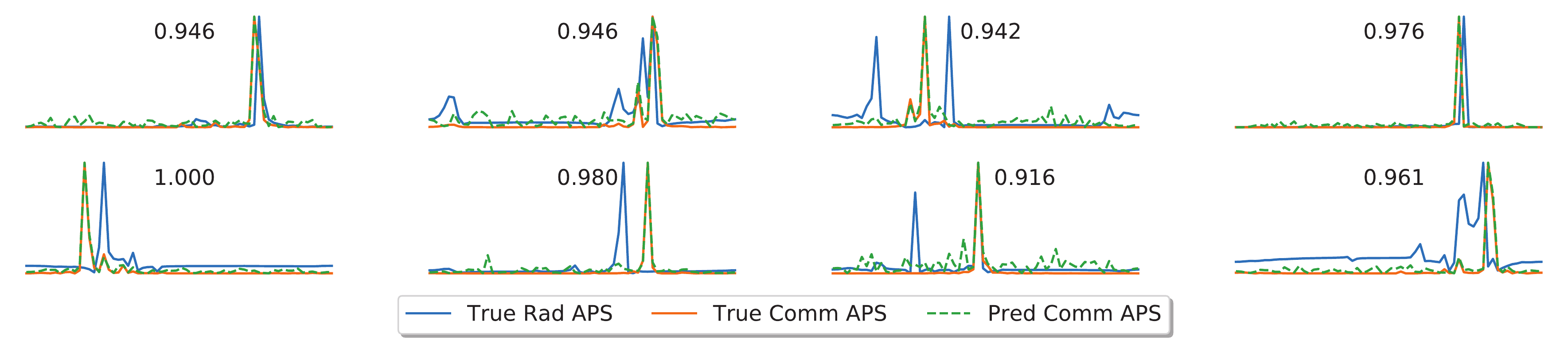}
	\caption{Examples of resulted communication APS from R2C covariance column prediction where the samples are randomly chosen from the test dataset. Column prediction captures the angle shift information and the resulted communication APS have high similarity with the true ones (shown on each subplot).} 
	\label{Toep_pred_exmps}
\end{figure*}

The similarity metric \cite{anum2020passive} compares two APS's $\bd_1$ and $\bd_2$ within a given window $L$, i.e., $S_{1\rightarrow 2}(L)=\frac{\sum_{i\in \mathcal{I}_1}\bd_2[i]}{\sum_{i\in \mathcal{I}_2}\bd_2[i]}$, where $\mathcal{I}_1$ ($\mathcal{I}_2$) contains indices of $L$ largest components of $\bd_1$ ($\bd_2$). This similarity metric rather than the MSE is used to evaluate the prediction accuracy, because the angles corresponding to the largest components of the APS are of particular interest for the beam search process.  
\begin{figure}
	\centering
	\subfloat[Good APS prediction]{%
		\label{good_aps_pred}
		\includegraphics[width=0.24\textwidth]{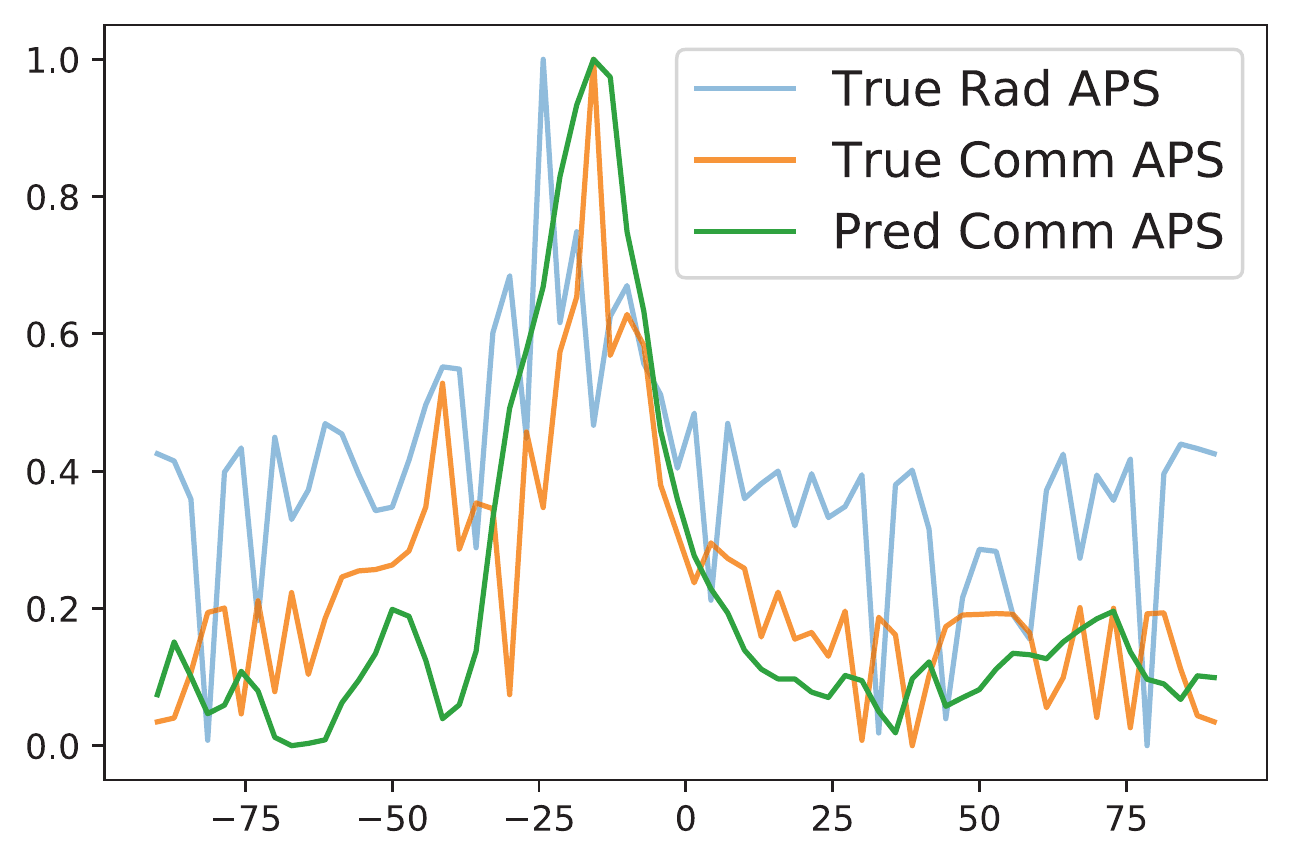}}
	\subfloat[Bad APS prediction]{%
		\label{bad_aps_pred}
		\includegraphics[width=0.24\textwidth]{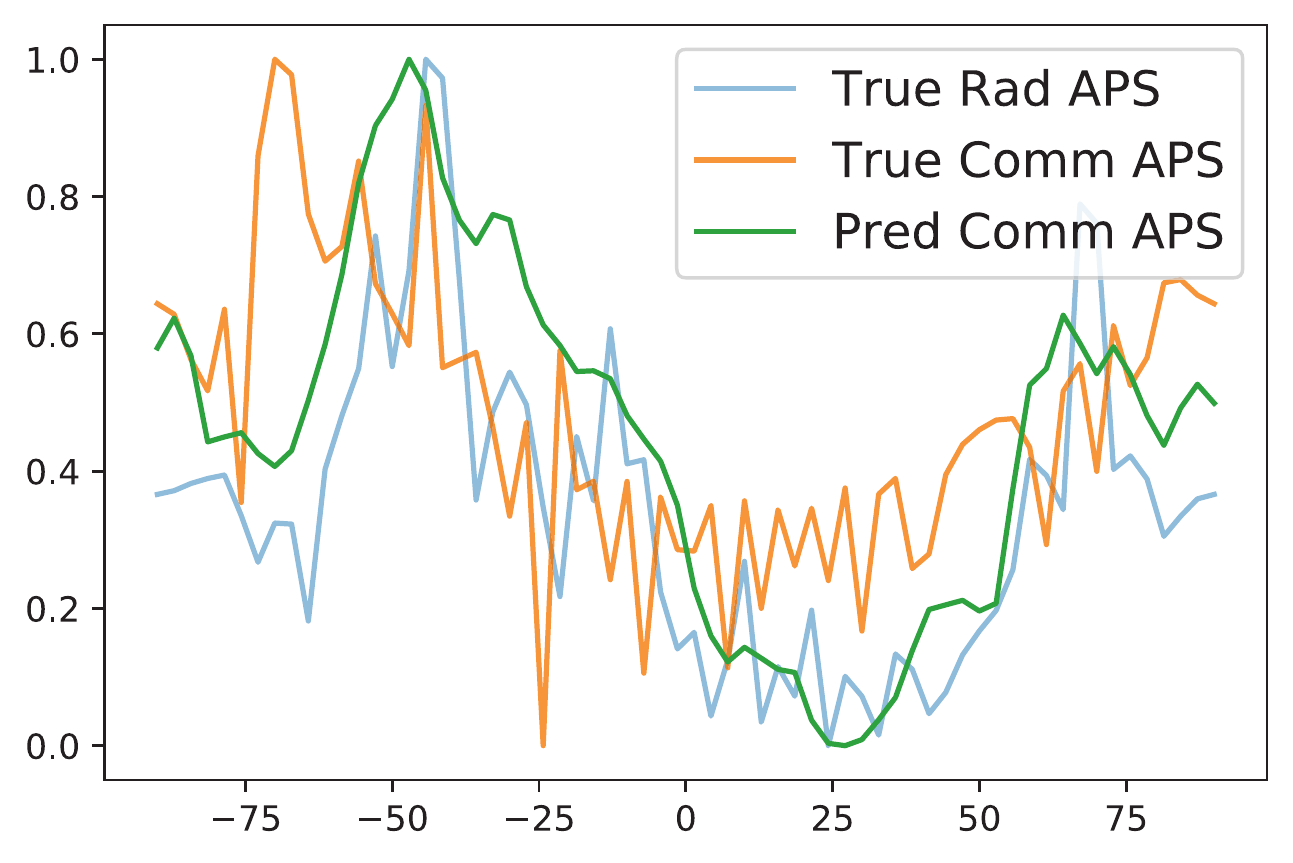}}
	\caption{Good and bad R2C APS (dB scale) prediction examples with the similarity value of (a) 0.9958 and (b) 0.7290 between the predicted and true communication APS.}
	\label{APS_pred_exmp}
\end{figure}
We evaluate the similarity between the estimated communication APS from the two prediction methods and the true communication APS. As an example shown in Fig.~\ref{APS_pred_exmp} with $L$ set to 5, higher similarity (Fig.~\ref{good_aps_pred}) indicates that the predicted APS and the true APS align better. While the APS prediction method cannot always capture the mismatching in the angle shift between the radar and communication APS (Fig.~\ref{bad_aps_pred}), covariance column predictions capture the mismatching and result in accurate communication APS estimations that align with the true APS, as shown in Fig.~\ref{Toep_pred_exmps}, where all the samples are randomly selected from the test dataset. 

\begin{figure}
	\centering
	\includegraphics[width=0.35\textwidth]{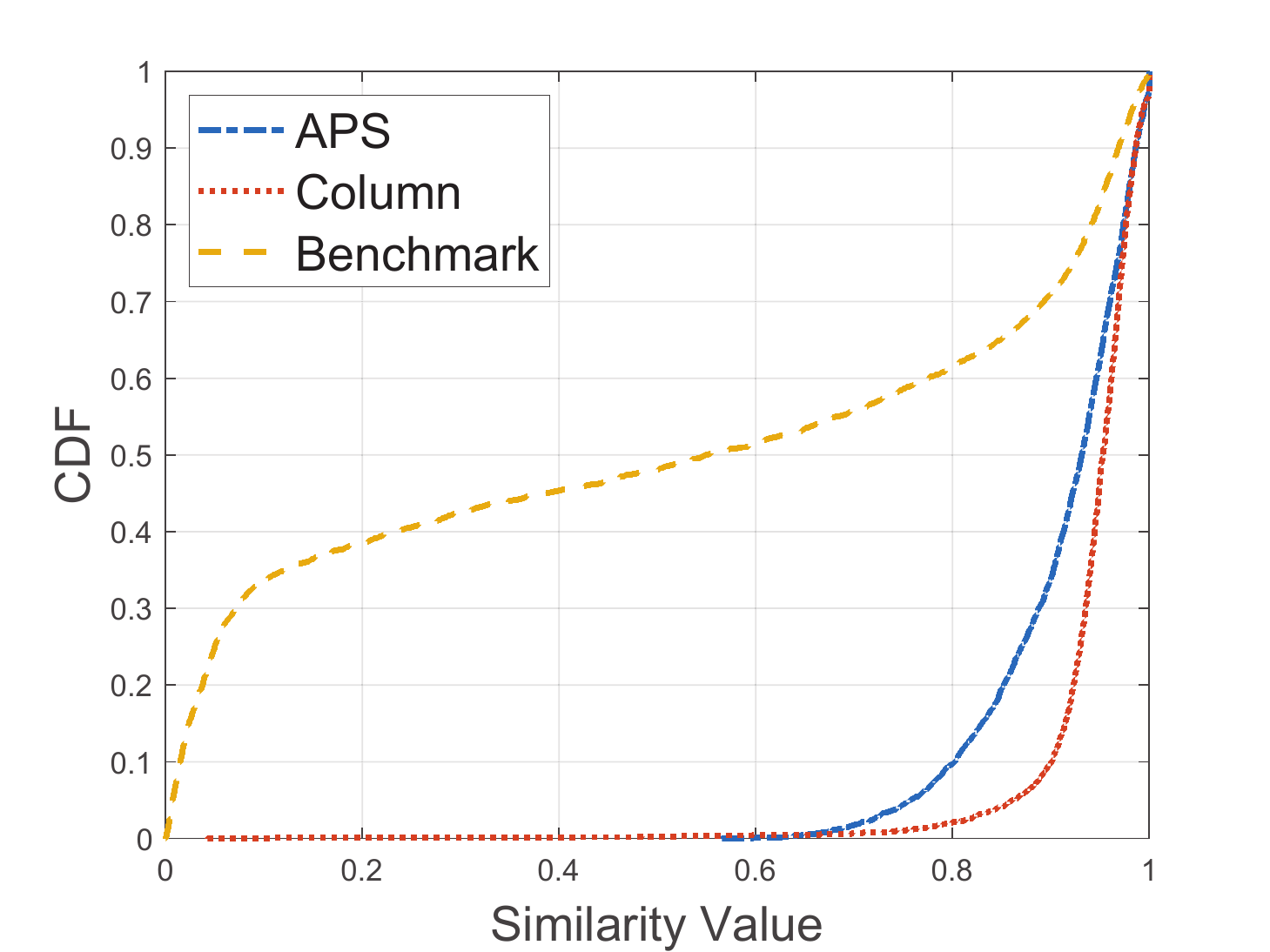}
	\caption{CDF of the similarity between the predicted communication APS and the true APS for the test dataset.} 
	\label{Simi_Perform}
\end{figure}

Fig. \ref{Simi_Perform} shows the cumulative distribution function (CDF) of similarity values ($L=5$) on the test dataset for the two proposed methods, with the similarity between the original radar APS and the true communication APS as the benchmark. The benchmark shows that the radar and the communication APS do not necessarily have high similarity, while our proposed APS prediction improves the similarity performance where the values are generally $\geq 0.6$. The covariance column prediction method outperforms the APS prediction method as high similarity occurs with higher frequency, specifically, the 10th-percentile similarity value achieves 0.9. The results are reasonable since for APS predictions, the DNN mainly captures the magnitude properties for the prediction, while the covariance column contains more completed inner features of the covariance matrix including the phase shift information. 
\subsection{Rate Results} \label{sec_rate_results}
We conduct an experiment to study how the R2C predictions assist establishing an mmWave communication link. In this experiment, we consider a single stream transmission ($N_s=1$) with the transmit power of 24 dBm using 2048 subcarriers. The bandwidth for the transmission is 491.52 MHz. 
We adopt the codebook design based on 2-bit phase-shifters \cite{ali2017millimeter}, while the region of the interest is a $180\degree$ sector spanning the angles $[-\frac{\pi}{2},\frac{\pi}{2})$. Thus, the $n$th codeword in the TX is $\frac{1}{\sqrt{N_v}}\ba_v(\arcsin(\frac{2n-N_v-1}{N_v}))$ and the codeword in the RX is defined similarly. We use the rate metric \cite{anum2020passive} for rate calculation considering a highly dynamic channel with the coherence time of $4N_{RSU}T_{sym}N_v$ (s), where $T_{sym}=4.7667$ ($\mu$s) is the symbol period.                                                                                                                                                                                                

Exhaustive search and radar-only search are treated as the benchmarks to compare with the two methods exploiting DL. For exhaustive search, the training overhead is $N_{RSU}\times N_v$ OFDM blocks, and the optimal beam-pair is determined by which brings the largest absolute value of the received signals over all the sub-carriers. Radar-only search reduces the beam search size at RX from 64 to 12, which finds the best combiner from the top 12 beams whose angles are close to the reference angle corresponding to the largest entry of the radar APS. The beam search size of 12 is found heuristically to maximize the rate by minimizing the likelihood of missing the optimal beam and minimizing the overhead. Similarly, we use a search size of 12 for the APS prediction method, the only difference is the reference angle which corresponds to the largest component of the predicted communication APS instead of the original radar APS. As the covariance column prediction has the best similarity performance, the search size could be further reduced to 2 according to the estimated communication APS from the prediction. 

\begin{figure}
	\centering
	\includegraphics[width=0.35\textwidth]{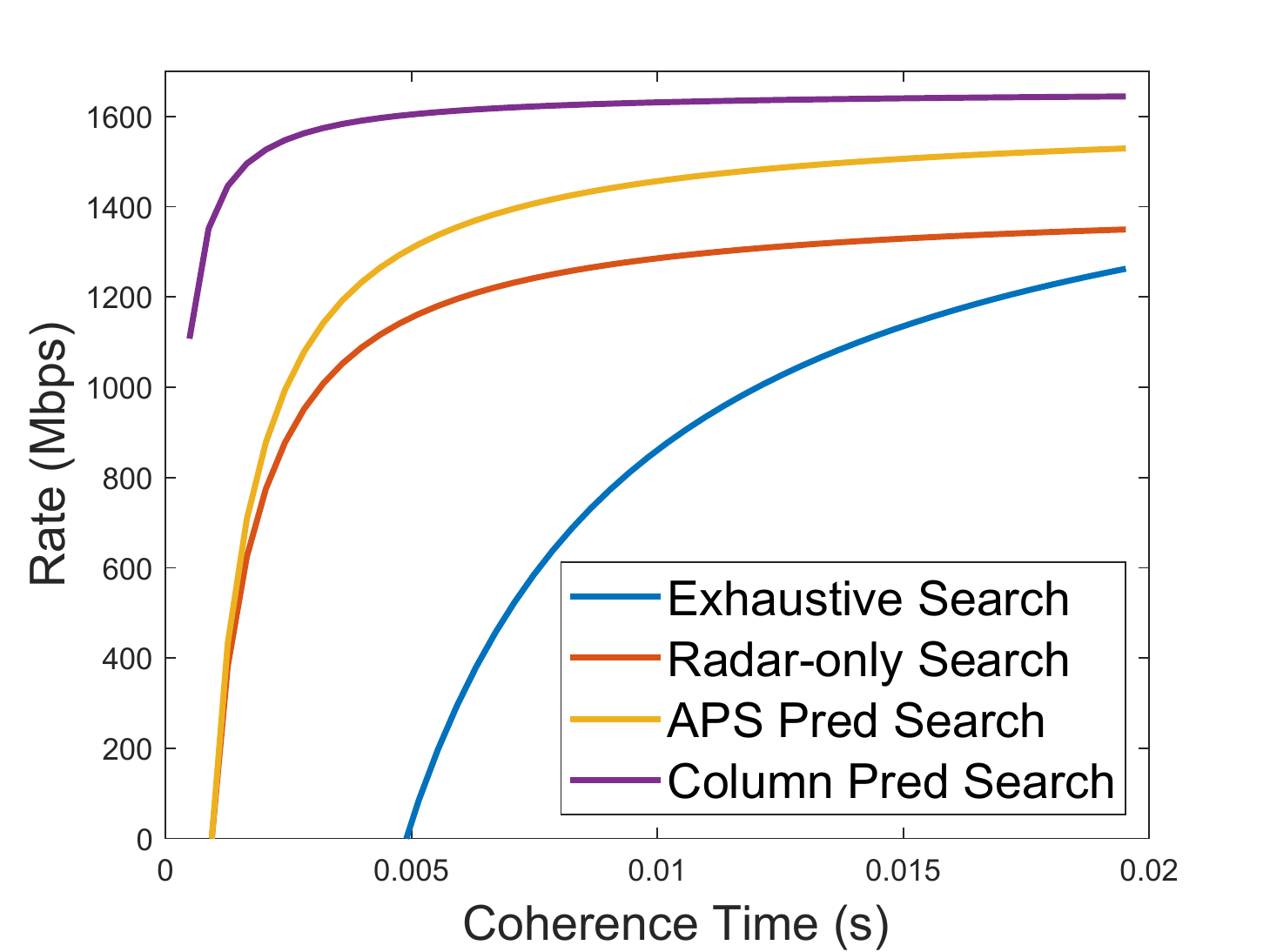}
	\caption{Rate results as a function of the coherence time. The beam search size is 12 for radar-only and APS prediction-based beam training and 2 for  covariance  prediction-based beamtraining.} 
	\label{Rate_Result}
\end{figure}

The rate results are fully based on the test dataset. The average rate of the samples in the test dataset is used for performance comparison for each method. As shown in Fig. \ref{Rate_Result}, exhaustive search achieves the lowest rate comparing with the other three methods, as it needs to allocate more resources to training blocks. The beam search based on R2C APS prediction requires similar training overhead as radar-only search due to the same beam search size, but later it achieves a rate of 1.450 Gbps, which is 200 Mbps (13.3\%) higher than the rate achieved by radar-only search. The covariance column prediction method outperforms all other three methods by achieving the final rate of 1.63 Gbps (21.9\% higher than radar-only search) with the lowest training overhead. 

Linking the rate results to the similarity performance, higher similarity in the APS from the DNN predictions implies more accurate R2C translation, which directly contributes to the efficient beam search. 

\section{Conclusions}
In this paper, we proposed two DNNs, one for R2C APS translation and another one for R2C covariance mapping. The predicted communication APS and covariance were used as prior information when establishing mmWave in a vehicular setting. The 10th-percentile value of the similarity between the estimated and the true communication APS achieves 0.8 and 0.9 for the APS prediction strategy and the covariance prediction method, respectively. Based on the high similarity, the beam training overhead has been significantly reduced, and the rate is increased by 13.3\% and 21.9\% using the APS and the covariance translation method when comparing with the rate results based on directly exploiting the radar APS without any mapping between the radar and communication channels.


\bibliographystyle{IEEEtran}
\bibliography{FullDuplexMmWave,Andrew}

\end{document}